\documentclass[a4paper,twocolumn,11.5pt]{article}
\usepackage[letterpaper,scale=0.875,right=1.3cm,top=2cm]{geometry} % Márgenes y demás
\input{CIENCIADESARROLLO.sty}  %%Llama el archivo de los estilos

{\title{
\vspace*{1cm}
\Large{ XY Traverse Stage for Automated Microscopy using Compliant Mechanism
\\}}}

\author[1,*]{\fontsize{10pt}{10pt}\selectfont \textbf{Venkata Sai Saran Grandhe}}
\author[1]{\textbf{Aditya Bandopadyay} }

\affil[1]{
\fontsize{8pt}{8pt}\selectfont Indian Institute of Technology Kharagpur, India}

\begin{document}

\twocolumn
[
\begin{@twocolumnfalse}
\maketitle
\begin{flushleft}
    % \textbf{Received:} XXXXX\\
    % \textbf{Accepted:} XXXXX\\
    % \textbf{How to cite:} XXXXX\\
    % \textbf{DOI:} XXXXX\\ 
\end{flushleft} 

\vspace*{0.3cm}

\renewcommand{\abstractname}{\textcolor{carnelian}{ABSTRACT}}
\begin{abstract}
\vspace*{0.5cm}
\fontsize{10pt}{10pt}\selectfont
\justify

This abstract provides an overview of research in precision engineering, with a focus on compliant mechanisms for XY stage positioning. The study explores existing methods, such as piezoelectric actuators and compliant mechanisms, as well as the work of renowned researchers in the field. It introduces two distinctive approaches for displacement reduction: topology optimization and flexure transmission mechanisms. The research aims to improve precision and reliability in automated microscopy and suggests a novel actuation procedure for the XY traversal stage using displacement reduction compliant mechanisms.

Compliant mechanisms are essential in this application due to their unique attributes, including high precision, reduced mechanical complexity. These mechanisms play a crucial role in XY stage positioning for automated microscopy, where accuracy and reliability are of paramount importance.

The study suggests potential future directions for validation, considering thermal effects, scalability, and decoupling challenges in XY stage applications, with the ultimate goal of advancing automated microscopy technology.

\keywordsEng{Compliant Mechanism, Topology Optimisation, Automated Microscopy}

\end{abstract}
\vspace*{0.3cm}

\renewcommand{\abstractname}{\textcolor{carnelian}{RESUMEN}}

% \begin{abstract}
% \vspace*{0.5cm}
% \fontsize{10pt}{10pt}\selectfont
% \justify

%   Se debe tener el resumen en ESPAÑOL, los cuales incluirán los objetivos principales de la investigación, alcance, metodología empleada, resultados principales y conclusiones. El resumen debe ser claro, coherente y sucinto, para lo cual se sugiere revisar y verificar datos, sintaxis, ortografía, no caer en erratas y no incluir ecuaciones, figuras, tablas ni referencias bibliográficas. El resumen es máximo de 150 palabras y debe reflejar fielmente el contenido del artículo.  Su redacción debe estar en tercera persona. Debe presentarse en un único párrafo.

% \keywordsSpan{Palabra 1, Palabra 2, Palabra 3.}
% \end{abstract}

\vspace{0.2cm}

\hspace*{0.7cm}

 %%% Espacio entre el abstract y las columnas

\vspace*{1cm} 
\end{@twocolumnfalse}]

%-----------------------------------------------

\section{INTRODUCTION}

\emph{Existing Methods}

In the field of precision engineering, there are a lot of ways by which the precise motion of the XY stage is achieved. In the “Design and Computational Optimisation of a Flexure-based XY Positioning Platform using FEA-based Response Surface Methodology” paper\cite{1}, it has been clearly explained how piezoelectric actuators were used. Alongside, compliant mechanisms serve as an important way in achieving the required motion without any mechanical joints.

Compliant mechanisms are a relatively new class of mechanisms that utilise compliance of their constituent elements to transmit motion and/or force. They can be designed for any desired input–output force–displacement characteristics, including specified volume/weight, stiffness, and natural frequency constraints. As flexure is permitted in these mechanisms, they can be readily integrated with unconventional actuation schemes, including thermal, electrostatic, piezoelectric, and shape-memory-alloy (SMA) actuators. Compliant mechanisms are single-piece flexible structures that deliver the desired motion by undergoing elastic deformation as opposed to rigid body motions of conventional mechanisms\cite{2}.

Renowned researchers in the field of compliant mechanisms, such as Howell and G.K. Ananthasuresh, have dedicated their efforts to fabricating a diverse array of compliant mechanisms tailored for various applications, spanning from displacement amplification to force amplification. Specifically, displacement-amplifying compliant mechanisms (DaCMs) have gained prominence for their ability to achieve substantial output displacement in relation to the input. Over the past decade, the design of DaCMs for actuator applications has garnered significant attention, as discussed in "Evaluation and Design of Displacement-Amplifying Compliant Mechanisms for Sensor Applications"\cite{3}. Notably, within sensor applications, Su and Yang, as well as Pedersen and Seshia, have harnessed force-amplifying compliant mechanisms (FaCMs) to enhance the sensitivity of resonant accelerometers, exemplified in the work of Su and Yang in "Two-Stage Compliant Micro-Leverage Mechanism Optimisation in a Resonant Accelerometer"\cite{4}.

The extensive advantages of compliant mechanisms, expounded upon in Howell's comprehensive work, encompass attributes such as high precision, reduced mechanical complexity, minimal backlash and friction, augmented repeatability, heightened reliability, and the capability to attain specific force-displacement characteristics. These attributes collectively underscore the indispensable role of compliant mechanisms in precision engineering applications, particularly within the realm of XY stage traversal for automated microscopy.

\emph{Different Approaches}

In the context of this paper, we delve into two distinctive approaches.

Firstly, we explore the utilisation of actuation methods involving servo motors or lead screws in tandem with displacement-reduction compliant mechanisms, achieved through topology optimisation inspired by Ananthasuresh's optimisation code, known as the YinSyn 2D Topology Optimisation code.

Secondly, we investigate the use of a flexure transmission-compliant mechanism, drawing inspiration from Hopkins JB's work on the design of parallel flexure systems through freedom and constraint topologies (FACT).

This paper also provides an in-depth introduction to topology optimisation for compliant mechanisms, showcasing the amalgamation of multiple mechanisms from FACT to construct a compliant stage. By combining these innovative techniques, this research contributes to the advancement of automated microscopy, ultimately facilitating even greater precision and reliability in specimen positioning and imaging.

\section{METHODOLOGY}

In order to achieve the displacement reduction, we followed two mechanisms,

\emph{Topology Optimization}

\paragraph{Spring Mass Lever Model}

Topology optimization is recognized as one of the systematic methods to design compliant mechanisms \cite{5}.This method, in its most popular implementation, operates on a fixed mesh of finite elements and defines a design variable, which is associated with each element in the mesh. The optimization algorithm determines the value of the design variables, which define the optimal topology of the mechanism. The design obtained from topology optimization is specific to the design domain, loading, and boundary conditions. The design variables are driven toward the optimal topology by the objective function and the constraints, which are specific to the problem. Objective function is a scalar quantity, which is a function of the design variables.Optimality criteria method \cite{6} is used in this paper to obtain new topology for the displacement reduction.

\begin{figure}[]
\centering
\includegraphics[height=5cm]{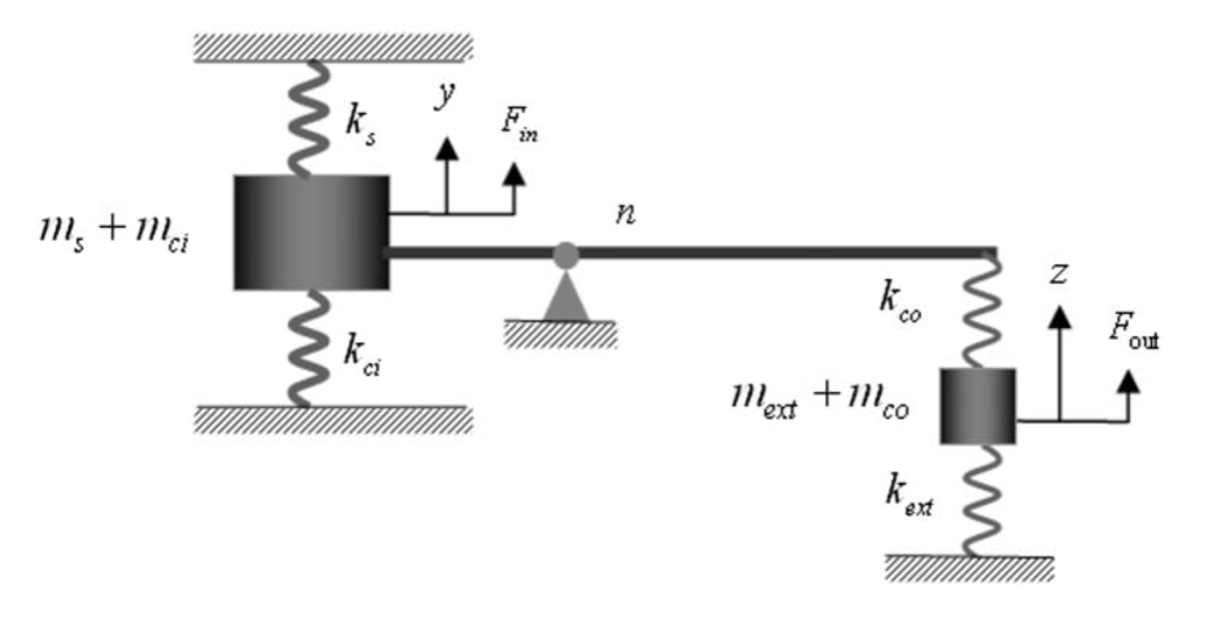}
\caption{Lumped Spring Mass Level Model}
\label{fig1}
\end{figure}

The existing DaCMs cannot be used with its input and output reversed. Unlike a rigid-body mechanism, compliant mechanism’s amplification and de- amplification factors are not simply reciprocals of each other. That is, the ratio of the output and input displacements is different when the input force is applied at either of the two points of interest. A lumped SML (spring mass lever) model that captures all the essential equations(\ref{eq1}, \ref{eq2}) is described to understand the above statement(Refer \ref{fig1}). 

\begin{equation}
\label{eq1}
  y = \frac{F_{out}(nk_{co}+ F_{in}(k_{co}+k_{ext})}{k_{co}(k_s+k_{ci})+k_{ext}(n^2k_{co}+k_{ci}+k_s)}
\end{equation}

\begin{equation}
\label{eq2}
  y = \frac{F_{in}(nk_{co}+ F_{out}(n^2k_{co}+k_{ci}+k_s)}{k_{co}(k_s+k_{ci})+k_{ext}(n^2k_{co}+k_{ci}+k_s)}
\end{equation}

\paragraph{Optimization Condition}

\begin{figure}[]
\centering
\includegraphics[height=7cm]{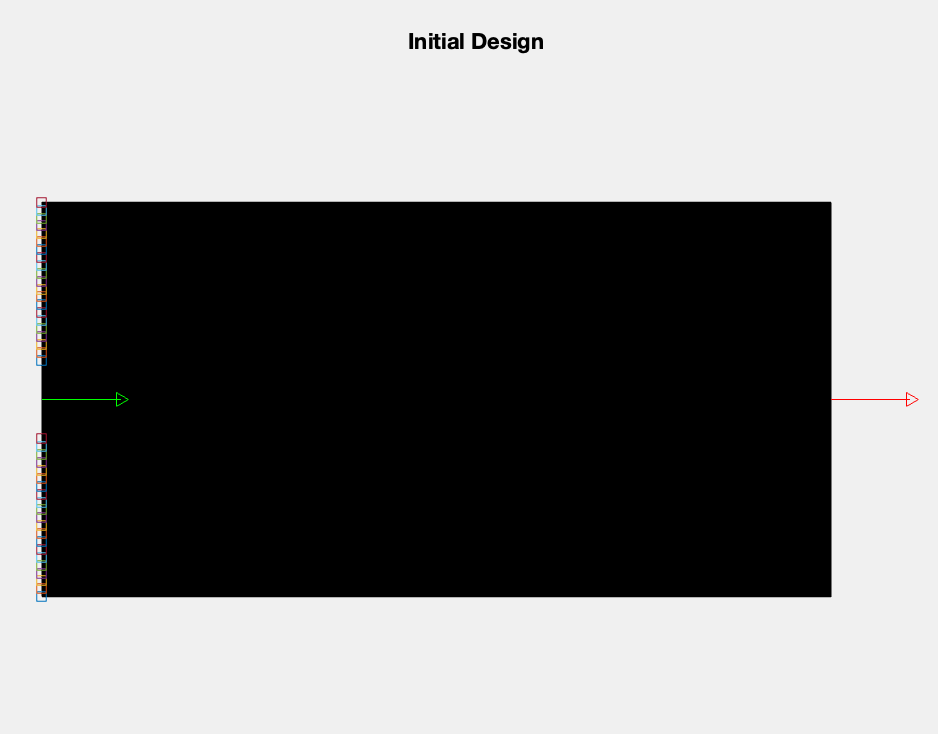}
\caption{Initial Domain Space}
\label{Initial}
\end{figure}

\begin{figure}[]
\centering
\includegraphics[height=7cm]{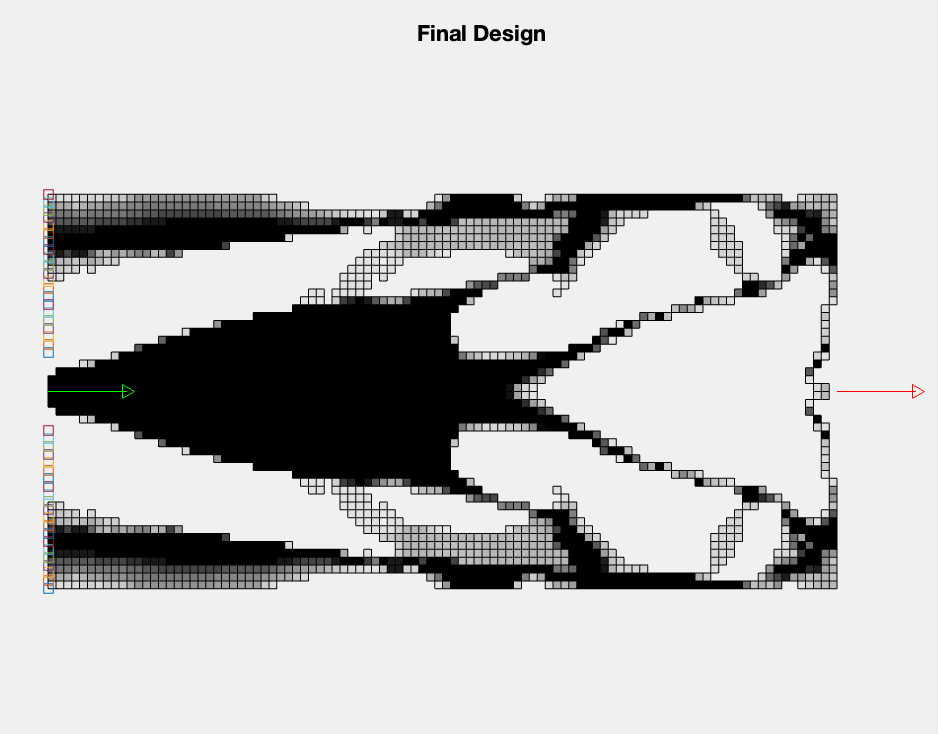}
\caption{Final Design Space after Optimisation}
\label{Final}
\end{figure}

For our mechanism, we need a displacement reduction objective function in Topology Optimisation analysis. Refer \ref{Initial}
and \ref{Final} for the input and output of the Topology Optimisation

\begin{equation}
\label{eq4}
  Minimise  \frac{u_{out}}{u_{in}}
\end{equation}

  Subject to

  Equilibrium equations

  Volume Constraints

  Bounds on the Variables

\begin{table}[!htbp]
\centering
\caption{Yin Syn Optimization Specifications}
\label{tab1}
\begin{tabular}{ll}

\hline
Finite Element Input & C0  \\\hline
Objective Function            & Compliant Mechanism \\\hline
Elements in X and Y directions & 100 mm, 50 mm  \\\hline
Input Port Specification & 15 mm \\\hline
Output Port Specification & 1 mm\\\hline
Material Properties & Plastic (PLA) \\\hline
\end{tabular}
\end{table}

\emph{FACT Mechanism}

\begin{figure}[]
\centering
\includegraphics[height=7cm]{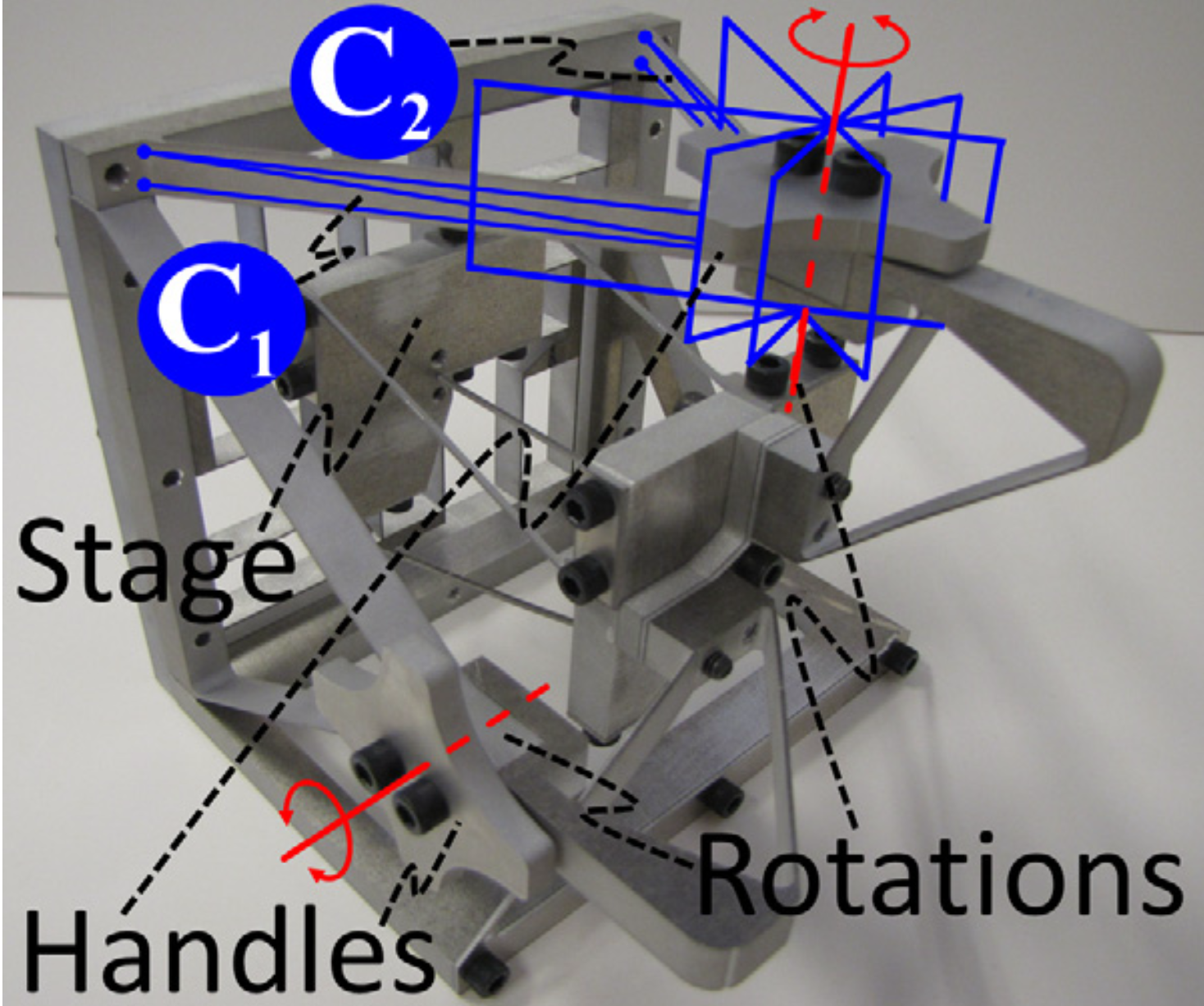}
\caption{Example of a complex multi-input flexure-based transmission mechanism synthesized using geometric shapes.}
\label{FACT Model}
\end{figure}

FACT utilises a comprehensive library of geometric shapes, like the intersecting blue planes shown in the figure \ref{FACT Model}, to enable designers to visualise every way a system may be synthesised with flexible elements for achieving any desired set of DOFs. In this way, designers may rapidly consider and compare numerous concepts without having to apply the time-consuming, rigorous mathematics of screw theory, which is already embodied by the geometric shapes.

Others have attempted to tackle the challenges of synthesising compliant transmission mechanisms. The three most common approaches for designing such mechanisms include the pseudo- rigid body model (PRBM), constraint-based design (CBD), and topological synthesis. 
Problems with each other are:

PRBM - Need Rigid body designing

CBD - Is only suitable for simple motions

Topological Synthesis - No common sense of designer is used. Only optimisation code is used

And, all together aren’t suitable for non-planer inputs.

\paragraph{Fundamental Spaces}

\begin{figure}[]
\centering
\includegraphics[height=7cm]{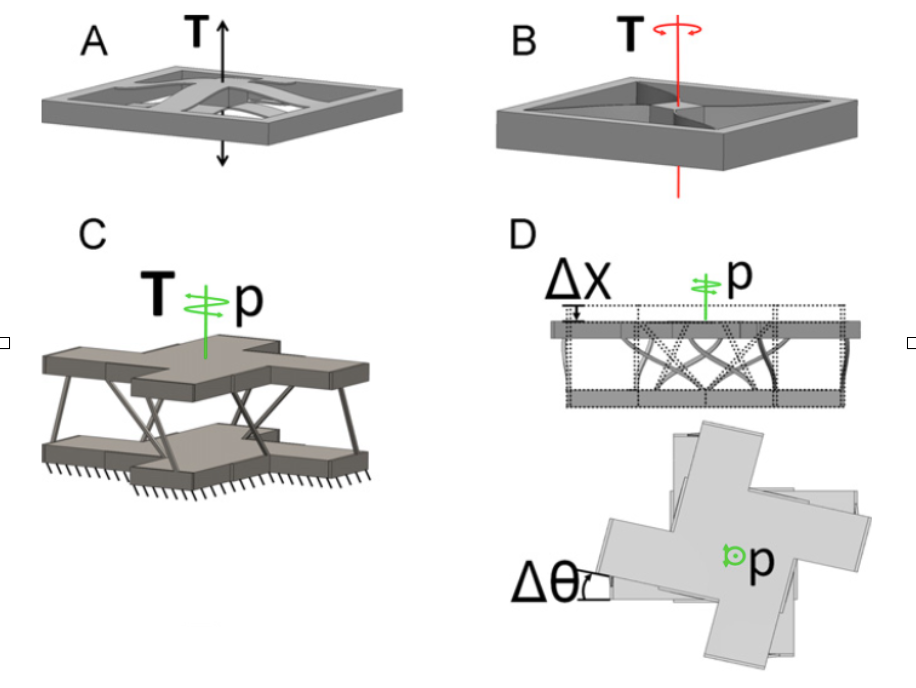}
\caption{Freedom Space for multiple components}
\label{Freedom Space}
\end{figure}

The fundamentals of FACT is dependent on 2 spaces:

\textbf{Freedom Space:} 
A system’s freedom space is a geometric shape that represents the complete kinematics of the system (i.e., every motion permitted by the system’s flexible constraints). According to JB Hopkins, these are the basic single DOF parallel, flexure systems. \ref{Freedom Space}

\begin{figure}[]
\centering
\includegraphics[height=7cm]{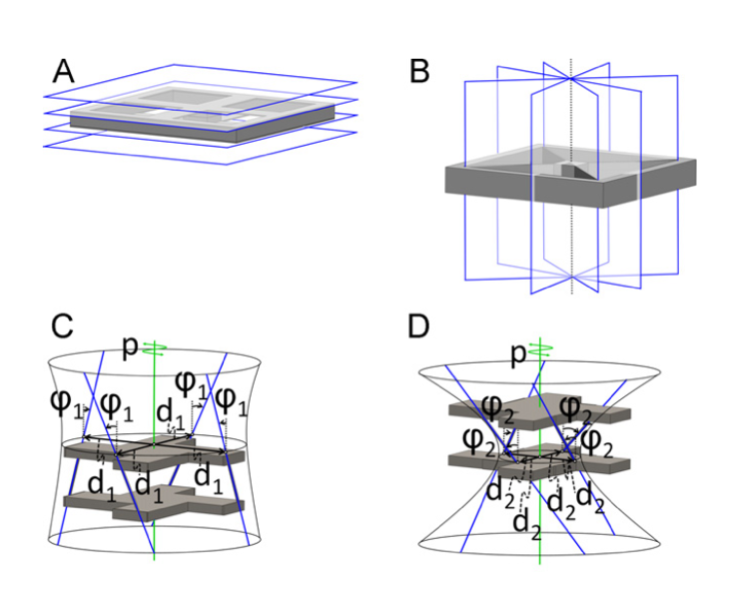}
\caption{Constraint Space for multiple components}
\label{Constriant Space}
\end{figure}

\textbf{Constraint Space:} 
Every freedom space uniquely links to a single constraint space. A system’s constraint space is a geometric shape that represents the regions wherein flexible constraints may be placed to permit the system’s DOFs. \ref{Constriant Space}

\paragraph{Transmission Train Mechanism}

\begin{figure}[]
\centering
\includegraphics[height=7cm]{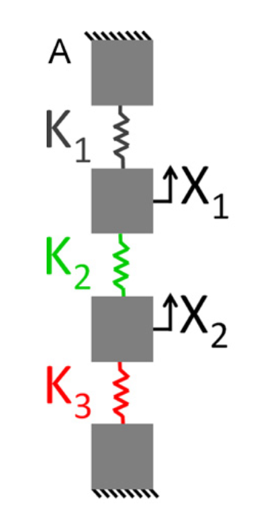}
\caption{Constraint Space for multiple components}
\label{Train}
\end{figure}

This section describes the mathematics necessary to calculate the transmission ratios of any flexure-based kinematic trans- mission train. Although the transmission ratio of a kinematic transmission train clearly depends on the topology of its constraints, the transmission ratio also depends on the train’s elastomechanics. This fact is most clearly demonstrated by considering the chain of linear springs shown in \ref{Train}. By applying the rules for calculating equivalent stiffnesses of linear springs we may determine that the transmission ratio of the chain of springs from \ref{Train} is 

\begin{equation}
\label{spring_train}
  \frac{X_2}{X_1} = K_3^{-1}.(K_3^{-1} + K_2^{-1})^{-1}
\end{equation}

And, the flexture Mechanism being similar to linear spring train mechanism, can be used for FACT with Screw Theory's Twist Wrench Stiffness Matrix. Screw theory’s twist–wrench stiffness matrix, TWSM \cite{7}, captures both of these characteristics. The TWSM of a parallel flexure system, \([K_{TW}]\), is a 6 × 6 matrix that links the flexure system’s motions to the forces/moments that cause these motions. Although this matrix is a type of stiffness matrix, the TWSM differs from conventional stiffness matrices in that it links twist vectors, T, to wrench vectors, W, rather than displacement vectors to force vectors according to \(W= [K_{TW}].T\) is \ref{Flexture_train} 

\begin{equation}
\label{Flexture_train}
  X_2 = [K_{TW}]_3^{-1}.([K_{TW}]_3^{-1}+ [K_{TW}]_2^{-1})^{-1}. T_1
\end{equation}

where, \(X_2\) is the linear displacement we need and \(T_1\) is the input twist given. 

\paragraph{CAD Model}

\begin{figure}[]
\centering
\includegraphics[height=7cm]{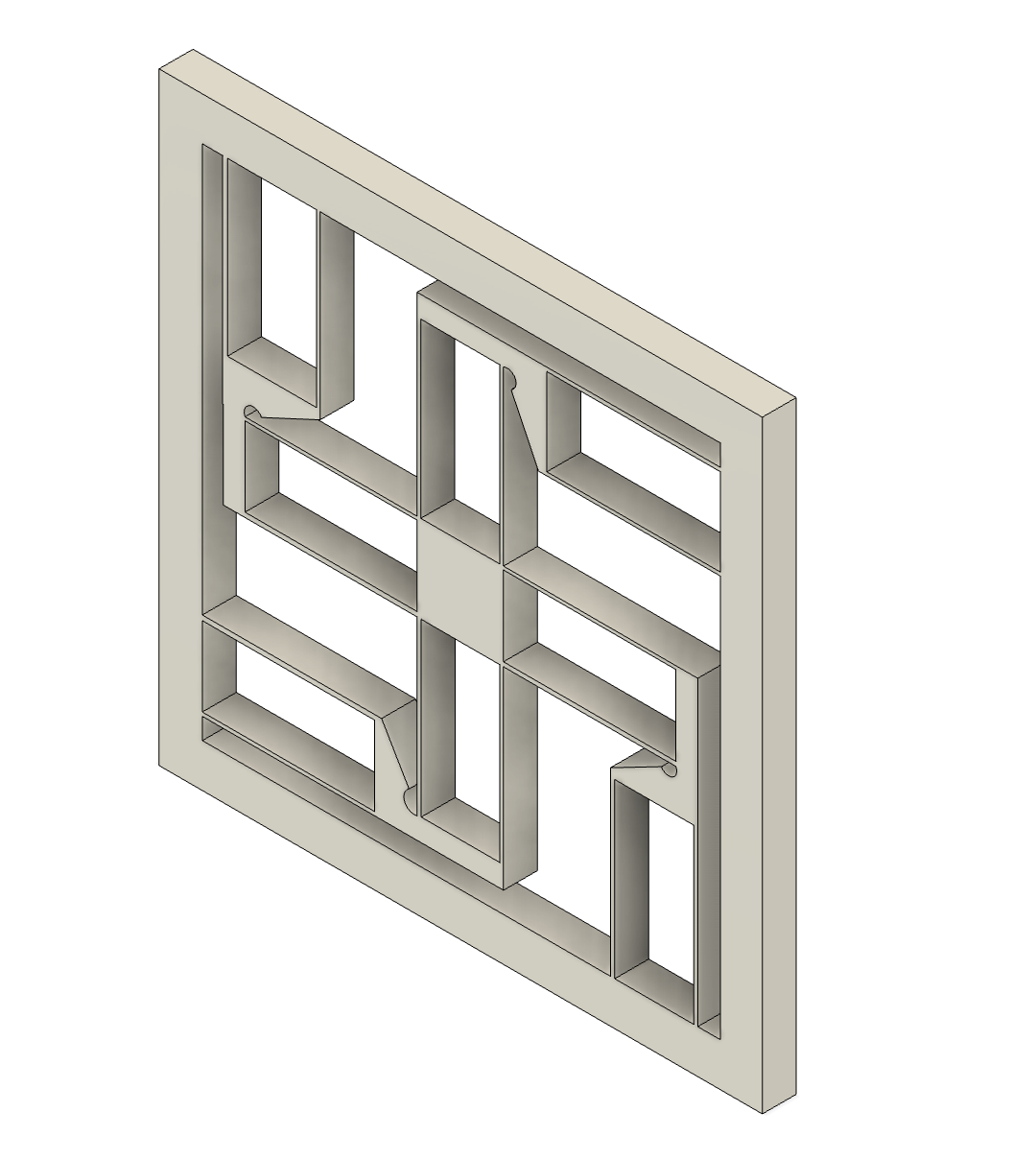}
\caption{Constraint Space for multiple components}
\label{CAD1}
\end{figure}

\begin{figure}[]
\centering
\includegraphics[height=7cm]{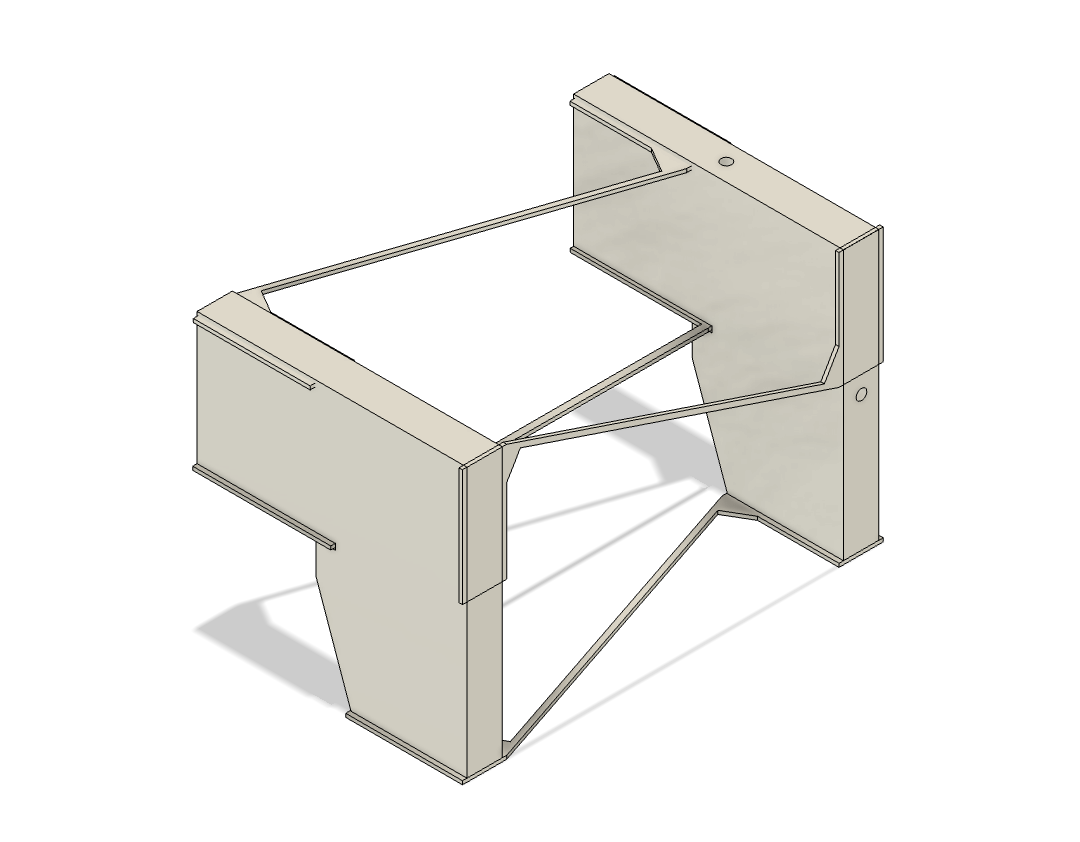}
\caption{Constraint Space for multiple components}
\label{CAD2}
\end{figure}

The compound mechanism is formed from transmission of \ref{CAD1} and \ref{CAD2} follows the equation \ref{Flexture_train}. The stage in \ref{CAD1} has a constraint space within the plane of flexture, whereas the flexture in \ref{CAD2} has a pitch. This is derived from Geometry. Four of the wire flexures lie on the surface of the circular hyperboloid 
\begin{equation}
\label{geometry}
   p=d_i·\tan{\theta_i}
\end{equation}
where \(p\) is the pitch of the screw and \(d_i, \theta_i\) are parameters defined in Figure D of \ref{Train}

\section{RESULTS}

\emph{FEA Results}

\paragraph{Flexture Mechanism}

The flexture assembly of figure \ref{CAD1} and \ref{CAD2} is used together for XY stage Traverse mechanism. The displacement is observed with respect to the actuation (Twist at the input side). The Finite Element Analysis is done using AutoDesk Fusion 360's simulation workspace. 

\begin{figure}[]
\centering
\includegraphics[height=5cm]{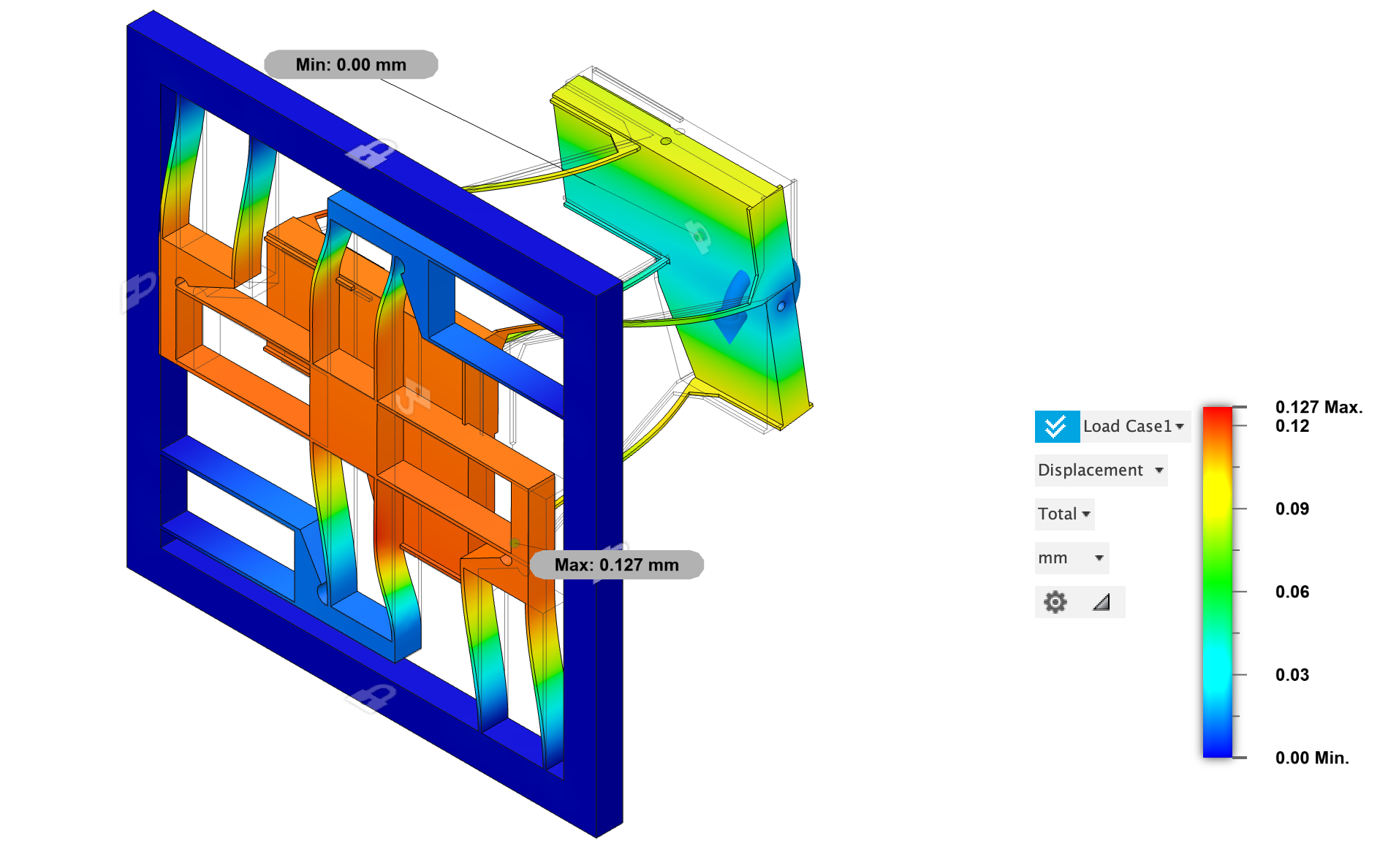}
\caption{FEA showing the adjusted deformation}
\label{FEA}
\end{figure}

NOTE: Adjusted Deformation in the structure is shown for easy visualisation purpose

The \ref{FEA} represents the adjusted deformation where the output displacement is \(13 \mu m\) for an input moment of 100 N/mm. This input moment corresponding to the input rotation of \(0.025 deg\). This corresponds to the resolution of \(520 \mu m/deg\).

Since the flexture componenet is symmetric, both the x and y axis resolutions are the same and that is proved from the simulation. 

The report of FEA is included in APPENDIX. It contains all the information of maximum stress created within the structure and minimum and maximum deformations.  

\paragraph{Topology Optimised Design}

\begin{figure}[]
\centering
\includegraphics[height=5cm]{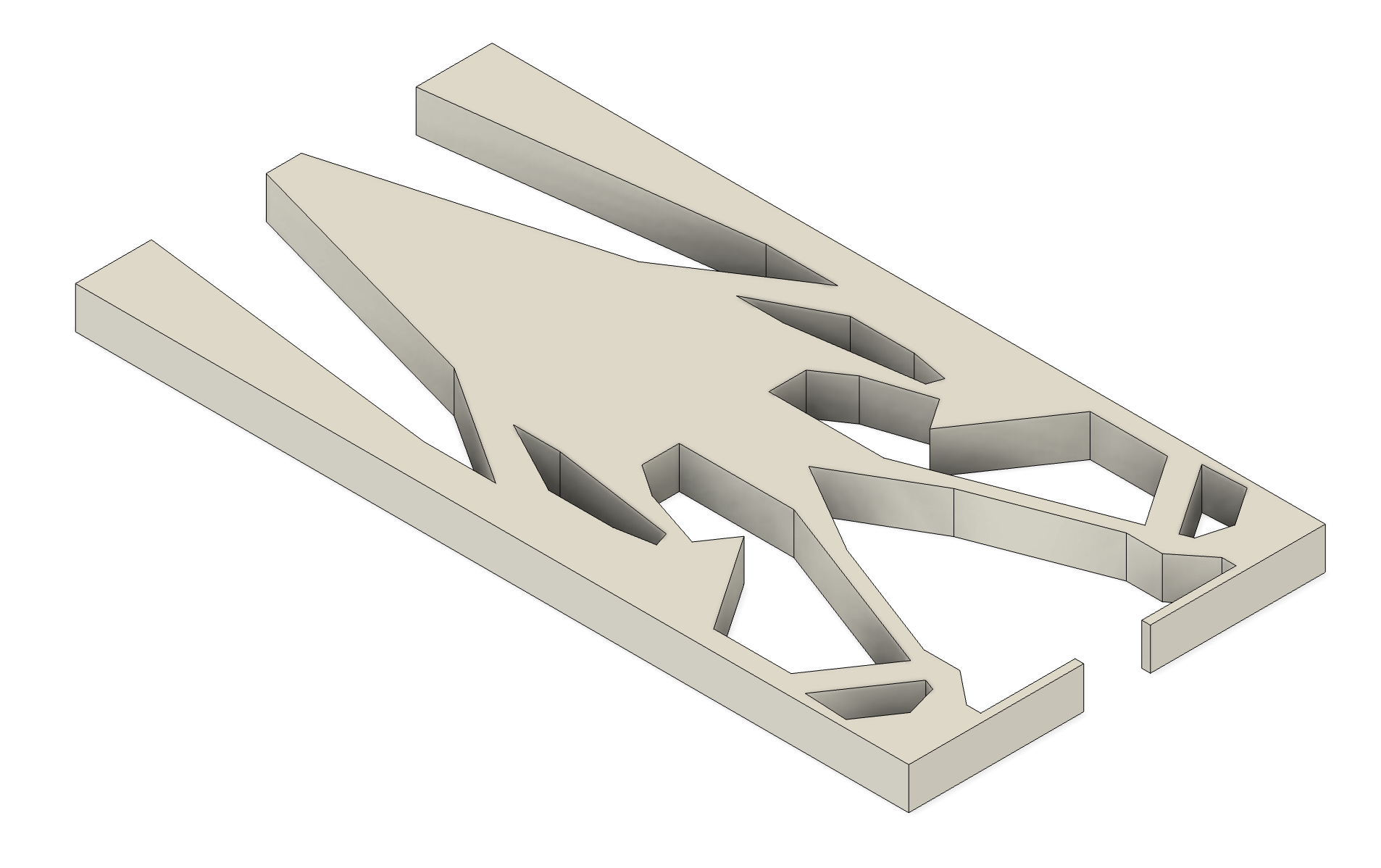}
\caption{CAD of the Topology Optimised Design}
\label{CAD_Topo}
\end{figure}

The Topology Optimised 2D design is modelled on a 3D CAD software. Refer \ref{CAD_Topo} for the CAD Model. 

For a 3D printed mechanism made of PLA, within the size of \(100 mm \times 50mm\), the best possible displacement reduction obtained is 5 times. This is determined keeping in mind the factor of safety not to be less than \(1.5\).

\begin{figure}[]
\centering
\includegraphics[height=5cm]{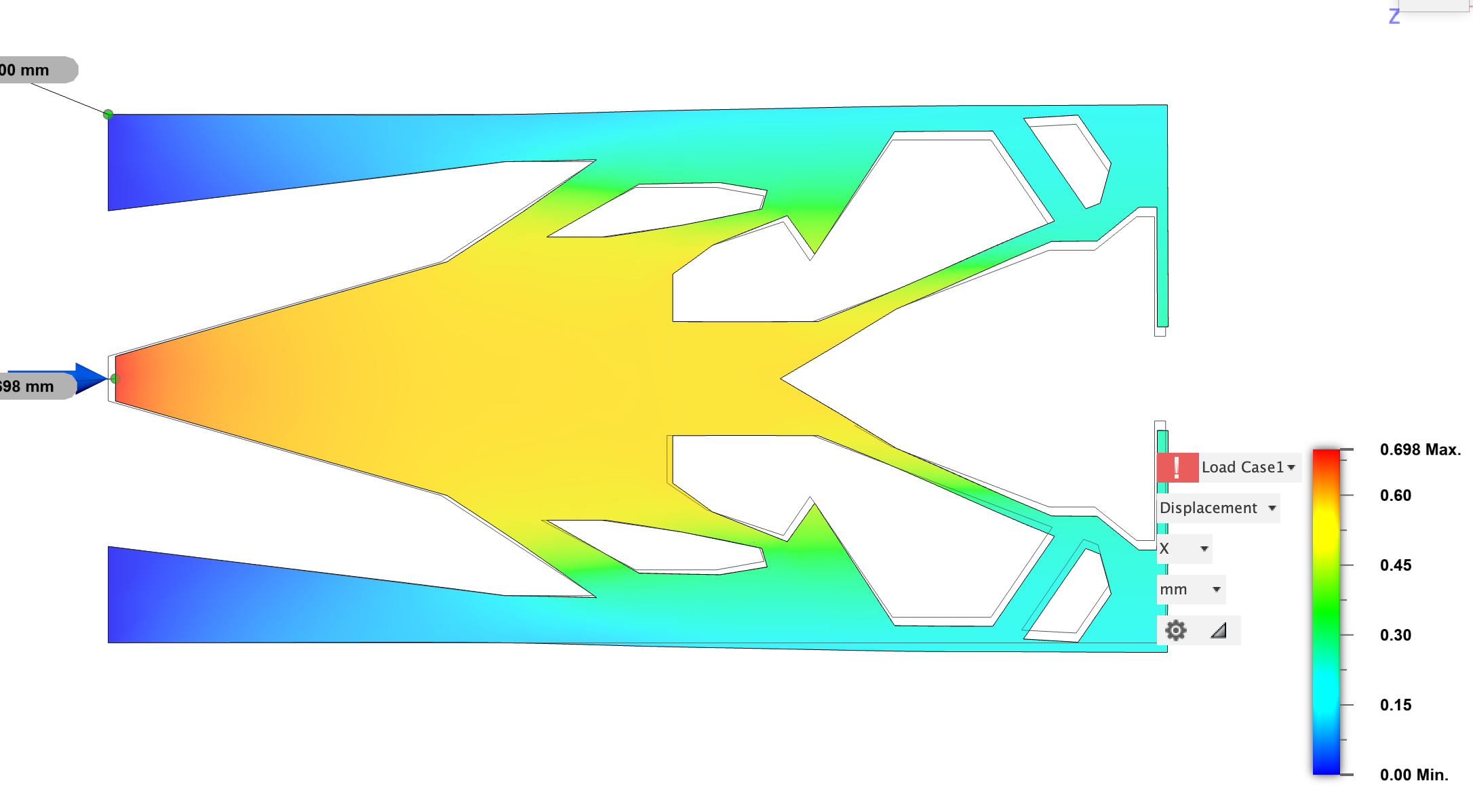}
\caption{CAD of the Topology Optimised Design}
\label{topo_FEM}
\end{figure}

The results of FEA, depicts the deformations at every point is shown in \ref{topo_FEM}

A detailed analysis and results of FEA are discussed in APPENDIX.

\emph{Comparision}

In comparing the two methodologies—FACT's Flexture mechanism and the Topology Optimized design—it becomes evident that each approach offers distinct advantages and trade-offs.

First, in terms of physical dimensions, the FACT mechanism tends to result in a larger structure compared to the topology-optimized design. The difference in size can have practical implications, particularly when spatial constraints are a crucial consideration in the application.

Second, the FACT mechanism introduces a slight parasitic error. While this error can be managed and compensated for, it remains a factor to carefully evaluate in precision-critical applications, impacting the overall accuracy of the system.

On the other hand, the topology-optimized design offers a more compact and space-efficient solution. However, it is important to note that this approach is primarily suitable for single-axis applications, limiting its applicability in certain contexts.

Furthermore, the topology-optimized design requires linear displacement as the input actuation, which necessitates the use of a lead screw to convert rotary input into linear motion. This transformation offers an opportunity for further optimization and control but introduces an additional mechanical element that needs to be considered.

In contrast, the FACT mechanism relies on direct rotary actuation. This simplifies the control process but may result in bulkier system designs, particularly when compactness is a priority.

Lastly, the topology-optimized design appears to exhibit a reduced degree of displacement reduction compared to the FACT mechanism.

\section{Discussion}

Both of these compliant mechanisms, employed in XY stage traversal for automated microscopy, have been effectively manufactured using 3D printing technology with PLA (Polylactic Acid). This method offers notable advantages, as it enables the creation of the entire compliant mechanism as a single, integrated piece, streamlining the manufacturing process. However, it's essential to recognize that these compliant mechanisms can also be crafted from metal through advanced machining techniques, such as Wire EDM (Electrical Discharge Machining). This alternative approach using metal offers distinct benefits, including heightened reliability and increased operational lifespan. Furthermore, metal-based compliant mechanisms exhibit a greater degree of resistance to vibrations, rendering them particularly suitable for applications in which the XY traverse stage may be subjected to external forces, as is often the case in micromachining processes that involve machine tool-induced impacts.

It's worth noting that for the specific context of automated XY traverse stages, the 3D printer design utilizing PLA proves to be an effective and practical solution. This approach strikes a balance between efficiency and performance, aligning well with the requirements of automated microscopy applications. However, the choice between 3D-printed PLA and metal-based compliant mechanisms should be made considering the specific demands of the application, weighing factors like reliability, durability, and environmental conditions that the XY stage may encounter during its operation.

\begin{figure}[]
\centering
\includegraphics[height=7cm]{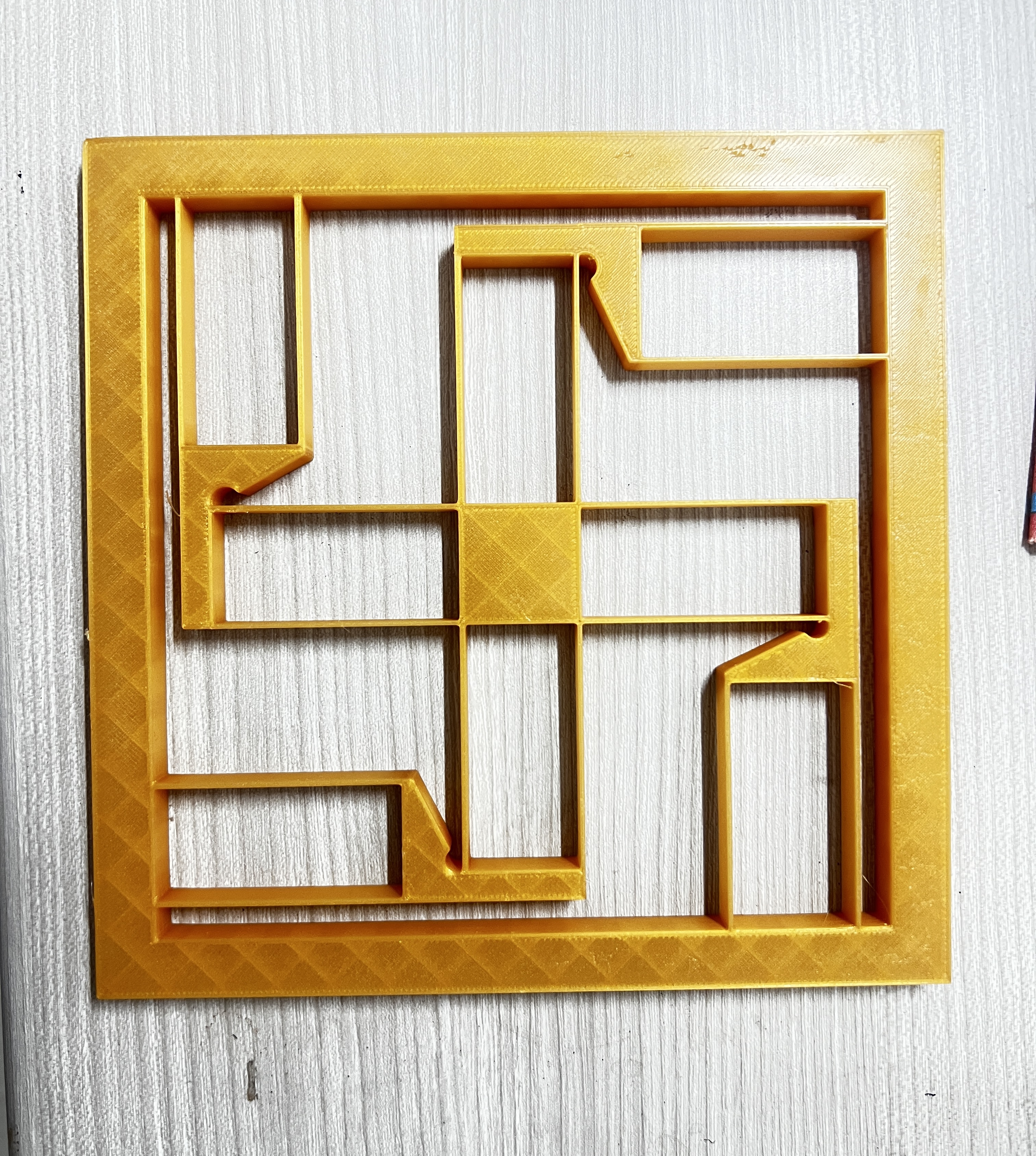}
\caption{3D printed Prototype - Underformed}
\label{Undeformed}
\end{figure}

\begin{figure}[]
\centering
\includegraphics[height=7cm]{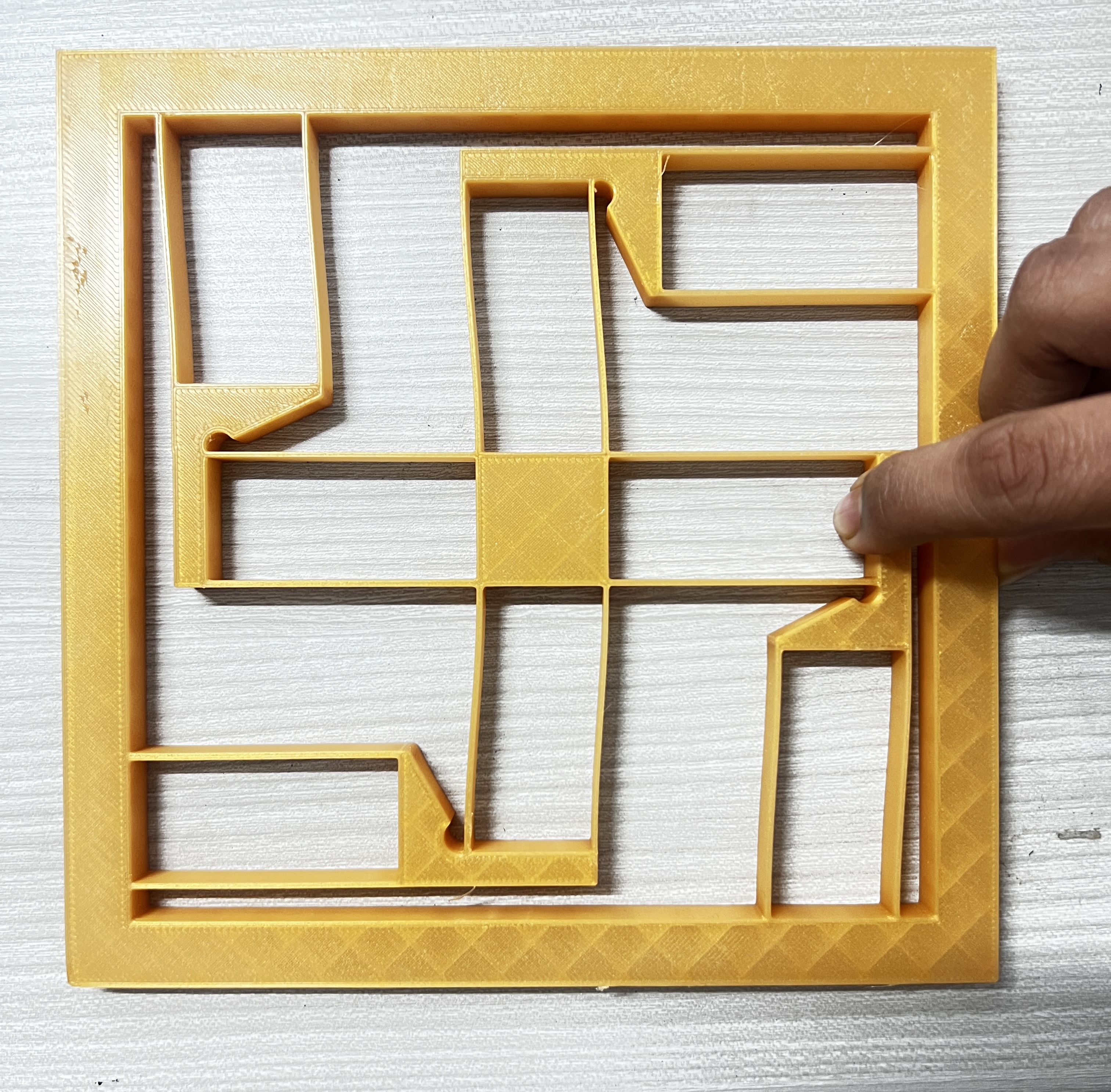}
\caption{3D printed Prototype - Deformed}
\label{Deformed}
\end{figure}

\begin{figure}[]
\centering
\includegraphics[height=7cm]{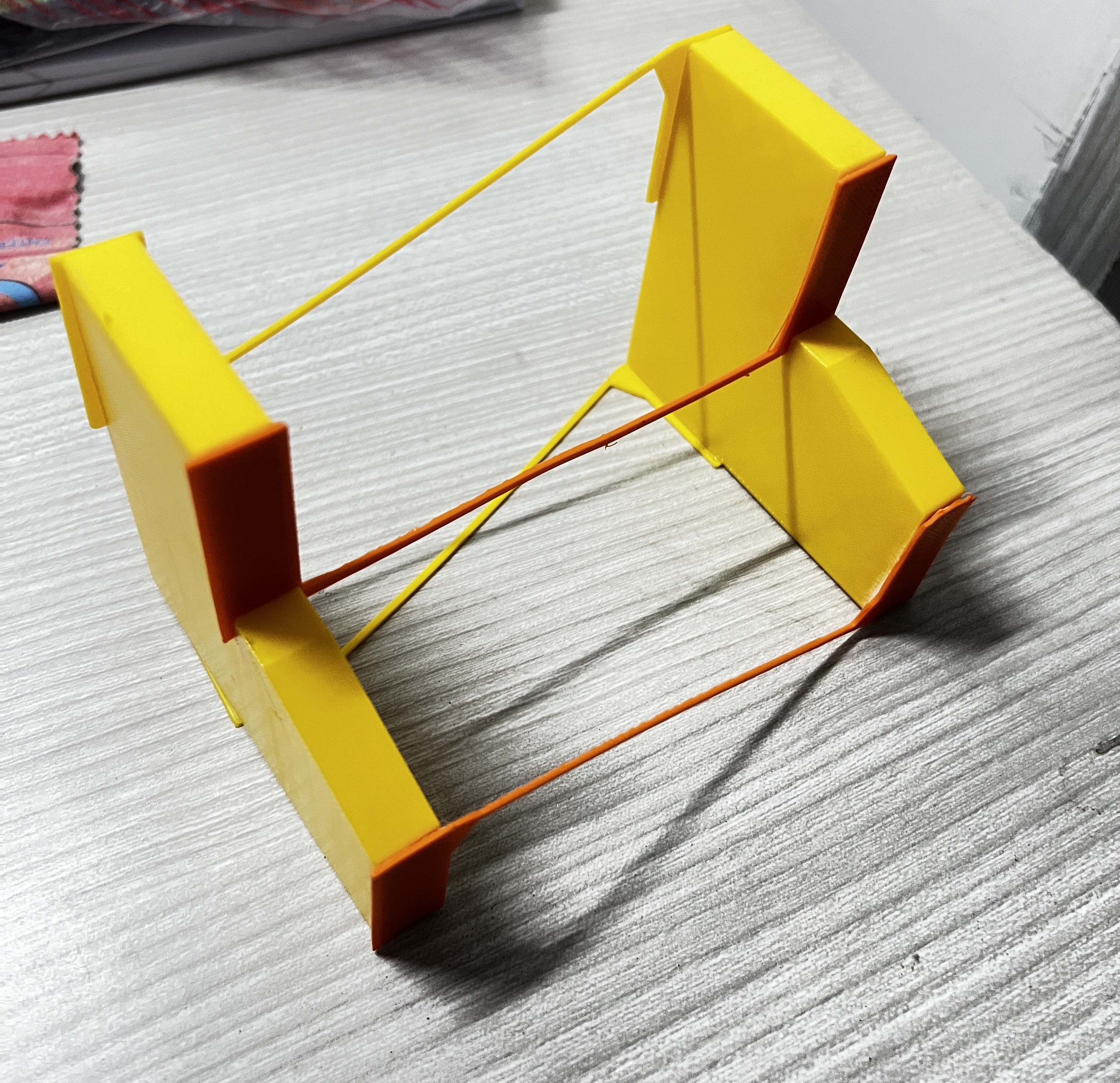}
\caption{3D printed Prototype - Untwisted}
\label{Untwisted}
\end{figure}

\begin{figure}[]
\centering
\includegraphics[height=7cm]{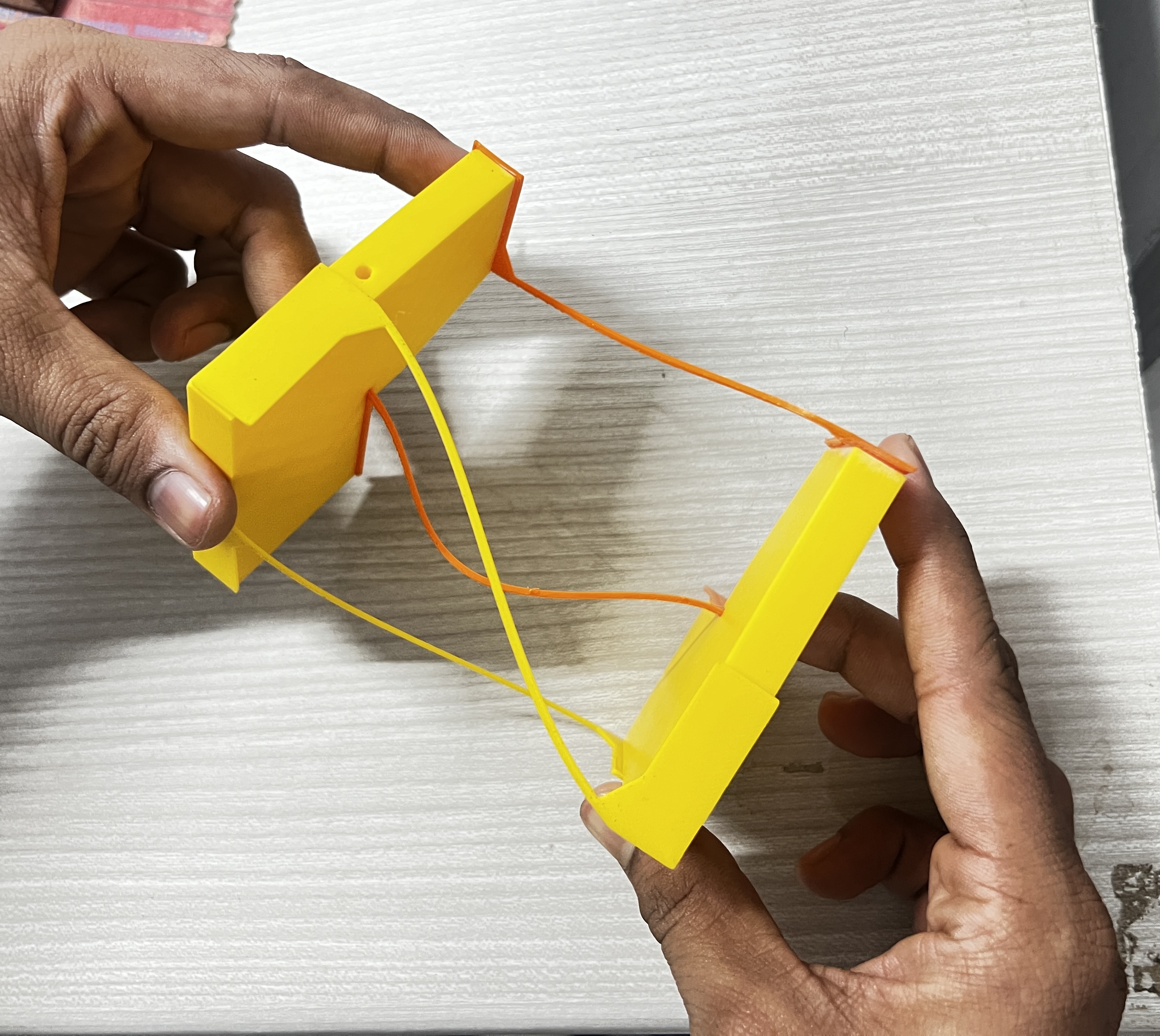}
\caption{3D printed Prototype - Twisted}
\label{Twisted}
\end{figure}

\section{Conclusion}

In this paper, we have demonstrated the utility of FACT mechanisms in synthesizing transmission mechanisms, while also delving into the concept of topology optimization, which has been effectively applied in practice. The results have been meticulously observed through numerical analysis, shedding light on the effectiveness of these compliant mechanisms.

Historically, displacement-amplifying compliant mechanisms have found application in actuator scenarios. In a novel extension of this concept, we have leveraged the same topology optimization methodology to design and fabricate an entirely new Displacement Reduction mechanism, expanding the horizons of compliant mechanism design.

Our exploration has enabled a comprehensive understanding of the capabilities of compliant mechanisms, spanning from micro-stage traversals to the deployment of folding solar panels in space. These mechanisms stand as crucial components in diverse applications, emphasizing their paramount importance in engineering and technology.

% \begin{thebibliography}{4}

% % No longer need the first \bibitem{1} entry

% \bibitem{2}
% Author, D., Author, E.F., and Author, G. (1995)
% \textit{Book Title}.
% Publisher Name, Publisher Address.

% \bibitem{1}
% @article{Brunner2017,
%   author = {Brunner, Philip and Therrien, René and Renard, Philippe and Simmons, Craig T. and Franssen, Harrie-Jan Hendricks},
%   title = {Advances in understanding river-groundwater interactions},
%   journal = {Reviews of Geophysics},
%   volume = {55},
%   number = {3},
%   pages = {818-854},
%   keywords = {river-aquifer interactions, modeling, streambed, remote sensing, geostatistics, parameter estimation},
%   doi = {https://doi.org/10.1002/2017RG000556},
%   url ={https://agupubs.onlinelibrary.wiley.com/doi/abs/10.1002/2017RG000556},
%   year = {2017}
% }

% \end{thebibliography}

\bibliographystyle{plain}
\bibliography{references} % Use the name of your .bib file here

\end{document}